\documentclass[epj]{svjour}

\usepackage{graphicx}
\usepackage{fancyhdr}

\def\mytitle{My title} 
\def\myauthors{My name}  
\def\mytype{My type of session}
\def\mysession{My session}

\def\mytitle{Radius stabilization in 5D SUGRA models on orbifold} 
\def\myauthors{Yutaka Sakamura}    
\def\mytype{Contributed Talk}    
\def\mysession{Theoretical Models}

\def\ignore#1{{}}

\newcommand{\alp}{\alpha}
\newcommand{\bt}{\beta}

\newcommand{\Gm}{\Gamma}
\newcommand{\dlt}{\delta}

\newcommand{\tht}{\theta}

\newcommand{\vtht}{\vartheta}
\newcommand{\kp}{\kappa}

\newcommand{\sgm}{\sigma}
\newcommand{\Sgm}{\Sigma}
\newcommand{\vph}{\varphi}
\newcommand{\omg}{\omega}
\newcommand{\Omg}{\Omega}

\newcommand{\be}{\begin{equation}}
\newcommand{\ee}{\end{equation}}
\newcommand{\bea}{\begin{eqnarray}}
\newcommand{\eea}{\end{eqnarray}}
\newcommand{\eql}{&=&}

\newcommand{\defa}{&\equiv&}

\newcommand{\simgt}{\stackrel{>}{{}_\sim}}

\newcommand{\tl}[1]{\tilde{#1}}
\newcommand{\bdm}[1]{{\mbox{\boldmath $#1$}}}

\newcommand{\diag}{{\rm diag}}
\newcommand{\der}{\partial}

\newcommand{\hc}{{\rm h.c.}}
\newcommand{\ie}{i.e.}

\newcommand{\brkt}[1]{\left( #1 \right)}
\newcommand{\brc}[1]{\left\{ #1 \right\}}
\newcommand{\sbk}[1]{\left[ #1 \right]}
\newcommand{\abs}[1]{\left| #1 \right|}

\renewcommand{\Re}{{\rm Re}\,}
\renewcommand{\Im}{{\rm Im}\,}

\newcommand{\cF}{{\cal F}}

\newcommand{\cL}{{\cal L}}

\newcommand{\cO}{{\cal O}}

\newcommand{\cV}{{\cal V}}

\begin{document}
\title{Radius stabilization in 5D SUGRA models on orbifold}
\subtitle{}
\author{Yutaka Sakamura\inst{1}
\thanks{\emph{Email:} sakamura@het.phys.sci.osaka-u.ac.jp}%
\thanks{\emph{} 
Based on a work in collaboration with Hiroyuki Abe~\cite{AS_moduli}}
}                     
%
%
\institute{Department of Physics, Osaka University, 
Toyonaka, Osaka 560-0043, Japan }
%
\date{}
\abstract{
We study a four-dimensional effective theory of the five-dimensional (5D) 
gauged supergravity with a universal hypermultiplet and perturbative 
superpotential terms at the orbifold fixed points. 
The class of models we consider includes the 5D heterotic M-theory 
and the supersymmetric Randall-Sundrum model as special limits 
of the gauging parameters. 
We analyse the vacuum structure of the models, especially 
the nature of the moduli stabilization, from the viewpoint of 
the effective theory. 
} 
\maketitle
%

\section{Introduction}
\label{intro}
The five-dimensional (5D) gauged supergravity compactified on 
an orbifold~$S^1/Z_2$ includes interesting models, such as 
the low-energy effective theory of the strongly 
coupled heterotic string theory~\cite{5D_M}, 
or the supersymmetric (SUSY) extension of 
the Randall-Sundrum (RS) models~\cite{RS}. 
Here we consider a certain class of models that is described as 
the 5D gauged supergravity with a universal hypermultiplet and 
superpotentials at the orbifold fixed points (boundaries). 
The hyperscalar manifold has an $SU(2,1)$ isometries and 
we gauge two directions among them by the graviphoton.\footnote{
In Ref.~\cite{AS_moduli}, we also consider 
gauging of one more independent direction.
}  
This theory is reduced to the above two models 
when we take certain limits of the gauging parameters. 
Our purpose here is to investigate the vacuum structures for 
this class of models~\cite{AS_moduli}.

\section{$\bdm{N=1}$ off-shell description of 5D action}
For our purpose the off-shell (superconformal) description of 
the 5D supergravity~\cite{KO} is useful 
because it enables us to treat the localized terms at the orbifold boundaries 
independently from the bulk action, and the isometries of the scalar manifold 
are {\it linearly} realized. 
Each 5D superconformal multiplet can be decomposed into 
$N=1$ multiplets. 
In our case, we have two compensator hypermultiplets~$(\Phi^1,\Phi^2)$, 
$(\Phi^3,\Phi^4)$, one physical hypermultiplet~$(\Phi^5,\Phi^6)$, 
and the graviphoton multiplet~$(V,\Sgm)$ besides the 5D Weyl multiplet.  
Here $\Phi^a$ ($a=1,2,\cdots,6$) and $\Sgm$ are $N=1$ chiral multiplets 
and $V$ is an $N=1$ vector multiplet. 
The 5D off-shell action is expressed in terms of 
these $N=1$ multiplets and an $N=1$ general multiplet~$V_E$ whose scalar 
component is $e_y^{\;\;4}$~\cite{N1_offshell}. 
Since $V_E$ has no kinetic term in the $N=1$ off-shell description, 
it can be integrated out and the 5D action is rewritten as~\cite{offshell_DR}
\bea
 \cL \eql -3e^{2\sgm}\int d^4\tht\;\cV\brc{d_a^{\;\;b}
 \bar{\Phi}^b\brkt{e^{-2(\tl{\alp}\cdot T)V}}^a_{\;\;c}\Phi^c}^{2/3} 
 \nonumber\\
 &&-e^{3\sgm}\sbk{\int d^2\tht\;\Phi^ad_a^{\;\;b}\rho_{bc}
 \brkt{\der_y-2(\tl{\alp}\cdot T)\Sgm}^c_{\;\;d}\Phi^d+\hc} \nonumber\\
 &&+\sum_{\vtht=0,\pi}\cL^\vtht\;\dlt(y-\vtht R)+\cdots, 
 \label{5Daction}
\eea
where $e^{\sgm}$ is the warp factor of the background metric, 
$d_a^{\;\;b}=\diag(\bdm{1}_4,-\bdm{1}_2)$, 
$\rho_{ab}=i\sgm_2\otimes\bdm{1}_3$, and 
$\cV\equiv -\der_y V+\Sgm+\bar{\Sgm}$ is a gauge-invariant quantity. 
The ellipsis denotes terms irrelevant 
to the following discussion. 
The boundary Lagrangians~$\cL^\vtht$ ($\vtht=0,\pi$) are 
written as 
\be
 \cL^\vtht = e^{3\sgm}\sbk{\int d^2\tht\;\Phi^2\Phi^3 P_\vtht
 \brkt{\frac{\Phi^5}{\Phi^3}}+\hc}, 
\ee
where $P_\vtht$ ($\vtht=0,\pi$) are the boundary superpotentials. 

The most general form of the gauging is parametrized by 
\be
 \tl{\alp}\cdot T\equiv \sum_{i=1}^8\tl{\alp}_i T^i, 
\ee
acting on $(\Phi^1,\Phi^3,\Phi^5)^t$ or $(\Phi^2,\Phi^4,\Phi^6)^t$, 
where $T^i$ ($i=1,2,\cdots,8$) are $3\times 3$ matrix-valued generators 
of $SU(2,1)$ shown in the Appendix of Ref.~\cite{AS_moduli}. 
The real coefficients~$\tl{\alp}_i$ determine the gauging direction. 
Now we consider a case that two independent isometries are gauged 
by the graviphoton, that is, $\tl{\alp}_i$ are parametrized by two 
parameters~$\alp$ and $\bt$ as 
\bea
 &&\tl{\alp}_3 = 2\bt, \;\;\;
 \tl{\alp}_6 = \alp, \;\;\;
 \tl{\alp}_8 = \alp+\bt, \nonumber\\
 &&\tl{\alp}_i = 0. \;\;\; (i\neq 3,6,8)
\eea
For simplicity we assume that the boundary superpotentials~$P_\vtht$ 
($\vtht=0,\pi$) consist of only constant and tadpole terms 
for the universal hypermultiplet, \ie, 
\be
 P_\vtht(Q) = w_\vtht^{(0)}+w_\vtht^{(1)}Q,  \label{bd_P}
\ee
where $w_\vtht^{(0)}$ and $w_\vtht^{(1)}$ are constants.

\section{4D effective action}
We can derive the four-dimensional (4D) effective action 
by the {\it off-shell dimensional reduction} proposed 
by Ref.~\cite{offshell_DR}, which are based on 
the $N=1$ superspace description~\cite{N1_offshell} of the 5D 
off-shell supergravity and developed in 
subsequent studies~\cite{sub_AS}. 
This method enables us to derive the 4D off-shell effective action 
directly from the 5D off-shell supergravity action 
{\it keeping the $N=1$ off-shell structure}. 
Note that only even multiplets under the orbifold $Z_2$-parity 
have zero-modes that appear in the effective theory. 
In our model the $Z_2$-even multiplets are $\Sgm$, $\Phi^2$, 
$\Phi^3$ and $\Phi^5$, and they appear in the 5D action only through 
the combinations of $\Sgm$, $\Phi^2\Phi^3$ and $\Phi^5/\Phi^3$, 
which have respectively the radion multiplet~$T$, 
the 4D chiral compensator~$\phi$ and the matter multiplet~$H$ 
as the zero-modes. 
Following the procedure of Ref.~\cite{offshell_DR}, we obtain 
the 4D effective action as\footnote{
We take the unit of the 4D Planck mass, \ie, $M_{\rm Pl}=1$. 
} 
\be
 S_{\rm eff} = -3\int d^4\tht\;\abs{\phi}^2e^{-K/3}
 +\brc{\int d^2\tht\;\phi^3 W+\hc}, 
\ee
where the K\"{a}hler potential~$K$ and the superpotential~$W$ are 
given by 
\bea
 K \eql -3\ln\brc{\frac{1}{2\alp}\cF(S_0,S_\pi)}, \nonumber\\
 W \eql e^{-\frac{3}{2}qS_0}(a_0+b_0 S_0)
 -e^{-\frac{3}{2}qS_\pi}(a_\pi+b_\pi S_\pi).  \label{KW_expr}
\eea
Here $q\equiv \bt/\alp$ and 
\be
 \cF(S_0,S_\pi) \equiv q^{-\frac{4}{3}}\brc{\Gm\brkt{\frac{4}{3},q\Re S_0}
 -\Gm\brkt{\frac{4}{3},q\Re S_\pi}}, 
\ee
where $\Gm(a,x)\equiv \int_x^\infty dt\;t^{a-1}e^{-t}$ is 
the imcomplete gamma function. 
The chiral multiplets~$S_\vtht$ ($\vtht=0,\pi$) are defined 
from $H$ and $T$ as 
\be
 S_0 \equiv \frac{1-H}{1+H}, \;\;\;\;\;
 S_\pi \equiv S_0+2\pi\alp T.  \label{def_Svtht}
\ee
The parameters~$a_\vtht$ and $b_\vtht$ in the superpotential~$W$ are given 
by linear combinations of the constants in the boundary 
superpotentials~(\ref{bd_P}) as
\bea
 a_\vtht \defa \frac{1}{8}\brkt{w_\vtht^{(0)}+w_\vtht^{(1)}}, \;\;\;\;\;
 b_\vtht \equiv \frac{1}{8}\brkt{w_\vtht^{(0)}-w_\vtht^{(1)}}. \nonumber\\
 &&(\vtht=0,\pi)
\eea

\ignore{
From the above 4D effective action, the scalar potential~$V$ is calculated as 
\bea
 V \eql \frac{-8\alp^3}{\cF(S_0,S_\pi)^2}\left[
 \brc{\frac{\cF(S_0,S_\pi)}{3}+\frac{(\Re S_0)^{4/3}e^{-q\Re S_0}}
 {1-3q\Re S_0}-\frac{(\Re S_\pi)^{4/3}e^{-q\Re S_\pi}}{1-3q\Re S_\pi}}^{-1}
 \right. \nonumber\\
 &&\hspace{23mm} 
 \times \abs{e^{-\frac{3}{2}qS_0}\frac{a_0-b_0\bar{S}_0}{1-3q\Re S_0}
 -e^{-\frac{3}{2}qS_\pi}\frac{a_\pi-b_\pi\bar{S}_\pi}{1-3q\Re S_\pi}}^2
 \nonumber\\
 &&\hspace{23mm}
 -4\left\{(\Re S_0)^{2/3}e^{-2q\Re S_0}
 \frac{\abs{b_0-\frac{3}{2}q(a_0+b_0S_0)}^2}{1-3q\Re S_0} \right. \nonumber\\
 &&\hspace{30mm} \left.\left.
 -(\Re S_\pi)^{2/3}e^{-2q\Re S_\pi} 
 \frac{\abs{b_\pi-\frac{3}{2}q(a_\pi+b_\pi S_\pi)}^2}
 {1-3q\Re S_\pi}\right\}\right]. 
\eea
}

\section{Heterotic M-theory limit}
In the limit $\bt\to 0$, Eq.(\ref{KW_expr}) becomes 
\bea
 K \eql -3\ln\sbk{\frac{3}{8\alp}\brc{(\Re S_\pi)^{4/3}
 -(\Re S_0)^{4/3}}}, \nonumber\\
 W \eql b_0\brc{C+S_0-rS_\pi}, 
\eea
where $C\equiv (a_0-a_\pi)/b_0$ and $r\equiv b_\pi/b_0$. 
The above K\"{a}hler potential reproduces the known result, \ie, 
the 4D effective K\"{a}hler potential of the heterotic 
M-theory~\cite{5D_M} when $\Re S_0 \gg\pi\alp\Re T$. 

The scalar potential~$V$ is calculated as 
\bea
 V \eql \brkt{\frac{8\alp}{3}}^3\abs{b_0}^2
 \left\{\frac{\abs{C-\bar{S}_0+r\bar{S}_\pi}^2}
 {\brc{(\Re S_\pi)^{4/3}-(\Re S_0)^{4/3}}^3} \right. 
 \nonumber\\
 &&\hspace{10mm}\left.
 -3\frac{\abs{r}^2(\Re S_\pi)^{2/3}-(\Re S_0)^{2/3}}
 {\brc{(\Re S_\pi)^{4/3}-(\Re S_0)^{4/3}}^2}\right\}. 
 \label{V_alp}
\eea
In this article we use the same symbols for the scalar fields as 
those for the chiral multiplets they belong to. 
From the SUSY preserving conditions:~$D_{S_0}W=D_{S_\pi}W=0$, 
we find a SUSY point, 
\bea
 &&(\Re S_0, \Re S_\pi) = \brkt{\frac{2\Re C}{r^4-1},
 \frac{2(\Re C)r^3}{r^4-1}}, \nonumber\\
 &&\Im C+\Im S_0-r\Im S_\pi = 0. 
\eea
From the second equation, we find a flat direction 
in the imaginary direction of $S_\vtht$ ($\vtht=0,\pi$). 
The superpotential~$W$ takes the nonzero value at this point.  
Thus the vacuum energy is negative, 
that is, the geometry is AdS${}_4$. 
By evaluating the second derivatives of the potential~(\ref{V_alp}), 
we can see that this SUSY point is a saddle point.  
Here we should note that SUSY points are always stable 
in a sense that they satisfy the Breitenlohner-Freedman 
bound~\cite{BFbound}. 
In Sect.~\ref{uplift}, we uplift 
the negative vacuum energy of the SUSY AdS$_4$ vacuum 
by a SUSY breaking vacuum energy in the hidden sector 
in order to obtain a SUSY breaking Minkowski vacuum, 
which is a candidate of our present universe.  
In general a SUSY saddle point remains to be a saddle point 
after the uplifting unless the uplifting potential is 
sufficiently steep, 
and it will not be stable any more. 
So we would like to look for a local minimum of the potential 
which is expected to be stable even after the uplifting.

\section{SUSY Randall-Sundrum limit}
In the limit~$\alp\to 0$, Eq.(\ref{KW_expr}) becomes 
\bea
 K \eql -3\ln\brkt{\frac{1-\abs{\Omg}^2}{2\bt}}
 -\ln(\Re S_0), \nonumber\\
 W \eql (a_0+b_0 S_0)-(a_\pi+b_\pi S_0)\Omg^3, 
\eea
where $\Omg\equiv e^{-\bt\pi T}$ is a warp factor superfield. 
The above $K$ reproduces the radion K\"{a}hler potential 
of the SUSY RS model~\cite{SUSY_RS}. 
Although $H$ is more conventional than $S_0$ for the SUSY RS model, 
we use the latter as a matter chiral multiplet because we will 
interpolate this model and the Heterotic M-theory limit 
discussed in the previous section. 
We can always translate $S_0$ to $H$ 
by the relation~(\ref{def_Svtht}). 

For simplicity, we assume in the following that the parameters satisfy 
a relation~$a_0b_\pi-b_0a_\pi = 0$, that is, 
\be
 (a_0,a_\pi) = c(b_0,b_\pi),  \label{rel_ab}
\ee
where $c$ is a constant. 
Then, from the SUSY conditions:~$D_{S_0}W=D_\Omg W=0$, we find a SUSY point, 
\be
 (S_0,\Omg) = (-c,r^{-1/3}), \label{SUSYpt_bt1}
\ee
for $\Re c<0$, and 
\be
 (S_0,\Omg) = \brkt{\bar{c},\abs{r}^{-4/3}\bar{r}^{1/3}}, 
 \label{SUSYpt_bt2}
\ee
for $\Re c>0$. 
Here $r\equiv b_\pi/b_0$. 
When $\Re c=0$, $\Omg$ is undetermined by the SUSY conditions 
and has a flat direction. 
At the SUSY point~(\ref{SUSYpt_bt1}) $W=0$ and thus the vacuum energy vanishes, 
resulting a local Minkowski minimum. 
This corresponds to the SUSY Minkowski vacuum discussed in Ref.~\cite{MO}, 
in which the boundary superpotentials~(\ref{bd_P}) consist of only the tadpole 
terms, \ie, $w_\vtht^{(0)}=0$ (or $c=-1$). 
On the other hand, the SUSY point~(\ref{SUSYpt_bt2}) is a saddle point 
and $W$ does not vanish there.  

The scalar potential~$V$ is calculated as 
\bea
 V \eql \frac{8\bt^3\abs{b_0}^2}{(1-\abs{\Omg}^2)^2\Re S_0}
 \left\{\frac{\abs{c-\bar{S}_0}^2\abs{1-r\Omg^3}^2}{1-\abs{\Omg}^2} \right.
 \nonumber\\
 &&\hspace{10mm}\left.
 -3\abs{c+S_0}^2(1-\abs{r}^2\abs{\Omg}^4)\right\}. 
 \label{V_bt}
\eea
Now we focus on the SUSY minimum~(\ref{SUSYpt_bt1}). 
We decompose the complex scalars into real ones as 
\be
 S_0 = s+i\sgm, \;\;\;\;\;
 \Omg = \omg e^{i\vph}. 
\ee
Evaluating the second derivatives of the scalar potential~(\ref{V_bt}), 
we can see that the four real scalars~$(s,\omg,\sgm,\vph)$ do not mix 
with each other. 
Then after normalizing them canonically, the mass eigenvalues are found as 
\bea
 m_s^2 \eql m_\omg^2 = 96\bt^3\abs{b_0}^2\abs{\Re c}
 \frac{\abs{r}^{4/3}(1+\abs{r}^{-2/3})^2}{1-\abs{r}^{-2/3}}, \nonumber\\
 m_\sgm^2 \eql m_\vph^2 = 48\bt^3\abs{b_0}^2\abs{\Re c}
 \frac{\abs{r}^{2/3}}{1-\abs{r}^{-2/3}}.  
 \label{moduli_mass}
\eea
We have assumed that $\Re c<0$ and $\abs{r}>1$.

\section{Interpolation between the two models}
\subsection{Vacuum structure}
In the vicinity of $\alp=0$, the K\"{a}hler and 
the superpotentials in (\ref{KW_expr}) are expressed as 
\bea
 K \eql -3\ln\left[\frac{1}{2\bt}\left\{1-\abs{\Omg}^2
 +\frac{1-\abs{\Omg}^2+\abs{\Omg}^2\ln\abs{\Omg}^2}{3q\Re S_0} 
 \right.\right.\nonumber\\
 &&\hspace{10mm}\left.\left.
 +\cO\brkt{\frac{1}{q^2(\Re S_0)^2}}\right\}\right]-\ln(\Re S_0), \nonumber\\
 W \eql b_0\brc{(c+S_0)(1-r\Omg^3)+\frac{2r}{q}\Omg^3\ln\Omg}. 
\eea
Here $q\abs{\Re c}$ is supposed to be 
large. 
From the SUSY conditions:~$D_{S_0}W=D_\Omg W=0$, 
we find a SUSY point as 
\bea
 S_0 \eql -c\left\{1-\frac{2}{3qc}\brkt{1-\frac{\ln r}{1-\abs{r}^{-2/3}}}\right.
 \nonumber\\ 
 &&\hspace{10mm}\left.
 +\cO\brkt{\frac{1}{q^2(\Re c)^2}}\right\}, \nonumber\\
 \Omg \eql r^{-1/3}\brc{1-\frac{\ln r}{9q\Re c}+\cO\brkt{\frac{1}{q^2(\Re c)^2}}}. 
 \label{SUSYpt_bt}
\eea
Since the SUSY point~(\ref{SUSYpt_bt1}) is a local minimum of the potential, 
this SUSY point is also a local minimum when $q\abs{\Re c}\gg 1$. 
Due to the correction from the SUSY RS limit, the superpotential~$W$ 
does not vanish at this point, 
\be
 W = -\frac{2b_0\ln r}{3q}\brc{1+\cO\brkt{\frac{1}{q^2(\Re c)^2}}}, 
\ee
and the vacuum energy is 
\bea
 V \eql -3e^{K}\abs{W}^2 \nonumber\\
 \eql -\frac{32\bt^3\abs{b_0\ln r}^2}{3q^2\abs{\Re c}(1-\abs{r}^{-2/3})^3}
 \brc{1+\cO\brkt{\frac{1}{q\Re c}}}. \nonumber\\
 \label{Venergy_bt}
\eea
Thus this is an AdS$_4$ SUSY vacuum. 

\subsection{Uplifting} \label{uplift}
From the AdS$_4$ SUSY vacuum such as (\ref{SUSYpt_bt}), we can obtain 
a SUSY breaking Minkowski minimum 
by introducing a sequestered SUSY-breaking sector 
just like the KKLT model~\cite{KKLT}. 
Following the KKLT model, the uplifting potential~$U$ is assumed 
as~\cite{choi} 
\bea
 U \eql \int d^4\tht\;(\bar{\phi}\phi)^n\kp\tht^2\bar{\tht}^2
 = \kp e^{nK/3} \nonumber\\
 \eql \frac{\kp(2\bt)^n}{(\Re S_0)^{n/3}(1-\abs{\Omg}^2)^n}
 \brc{1+\cO\brkt{\frac{1}{q\Re c}}}, \nonumber\\
\eea
where $\kp$ is a constant. 
The typical value of $n$ for the sequestered SUSY breaking source is 
given by $n=2$. 
The total scalar potential is then given by $V_{\rm tot}\equiv V+U$. 
If we choose $\kp$ as 
\be
 \kp = \frac{4(2\bt)^{3-n}\abs{b_0\ln r}^2}{3q^2\abs{\Re c}^{1-n/3}
 (1-\abs{r}^{-2/3})^{3-n}}\brc{1+\cO\brkt{\frac{1}{q\Re c}}}, 
\ee
then the minimum value of the total potential~$V_{\rm tot}$ vanishes, 
and we can obtain a SUSY breaking Minkowski vacuum. 
At this minimum, we can evaluate the order parameter of the SUSY breaking. 
Following Ref.~\cite{choi}, we define the anomaly/modulus ratio of SUSY breaking as
\be
 \alp_{A/M}\equiv \frac{1}{\ln(M_{\rm Pl}/m_{3/2})}\cdot
 \frac{F^\phi/\phi}{F^T/(T+\bar{T})}, 
\ee
where $F^\phi$ and $F^T$ are the F-terms of $\phi$ and $T$ respectively. 
Then we find that 
\bea
 \alp_{A/M} \eql \frac{q\Re c}{\ln(M_{\rm Pl}/m_{3/2})}
 \left\{\frac{6\ln\abs{r}}{n\ln r}\abs{r}^{2/3}
 (1+\abs{r}^{-2/3})^2 \right. \nonumber\\
 &&\hspace{10mm} \left.
 \times (1-\abs{r}^{-2/3}) 
 +\cO\brkt{\frac{1}{q\Re c}}\right\}. 
\eea
Since $\abs{r}=e^{3\pi\bt\Re T}$ from (\ref{SUSYpt_bt1}), 
we can see that $\abs{\alp_{A/M}}\gg 1$ unless $\bt$ is small.\footnote{ 
Note that the radius of the orbifold~$\Re T$ should be larger than 
the Planck length~$M_{\rm Pl}^{-1}=1$.
}
Thus the anomaly mediation tends to dominate in this model. 
However, for small values of $\bt$, the parameter~$\abs{r}$ is 
allowed to be of $\cO(1)$ and the modulus mediated contribution 
can be comparable to that of the anomaly mediation. 
For example, $\abs{\alp_{A/M}}\simeq 1$ when $n=2$, $r=2$, $q\Re c=-8$, 
$\ln(M_{\rm Pl}/m_{3/2})=4\pi^2$. 
In this case, the mirage mediation is realized. 
Finally note that the moduli masses, which are given by (\ref{moduli_mass}) 
at the leading of the $(q\Re c)^{-1}$-expansion, are much larger than 
the gravitino mass, 
\bea
 m_{3/2}^2 \eql e^K\abs{W}^2 \nonumber\\
 \eql \frac{32\bt^3\abs{b_0\ln r}^2}{9q^2\abs{\Re c}(1-\abs{r}^{-2/3})^3}
 \brc{1+\cO\brkt{\frac{1}{q\Re c}}}. \nonumber\\
\eea

\section{Summary}
We studied the 4D effective theory of the 5D gauged supergravity on 
an orbifold with a universal multiplet and boundary superpotentials. 
We analysed a class of models obtained by gauging two independent isometries 
on the scalar manifold. 
It includes the 5D heterotic M-theory and the SUSY RS model 
as special limits of the gauging parameters. 
We have investigated the vacuum structure of this class of models 
and the nature of moduli stabilization 
assuming perturbative superpotentials at the orbifold boundaries. 

In the heterotic M-theory limit, the SUSY point is a saddle point 
of the potential. 
In the SUSY RS limit, on the other hand, the SUSY point is a local minimum 
with vanishing vacuum energy when the parameters satisfy 
the relation~(\ref{rel_ab}) with $\Re c<0$. 
These SUSY points in the two different limits continuously transit 
to each other by changing the ratio~$q=\bt/\alp$~\cite{AS_moduli}. 
When $\abs{q\Re c}\gg 1$, there is a SUSY AdS$_4$ vacuum 
which is a good candidate for the KKLT-type uplifting. 
Thus we studied the uplifting of this vacuum and find that 
the mirage mediation can be realized for small values of $\bt$, 
while the effect of the anomaly mediation is dominant for $\bt\simgt\cO(1)$. 
The moduli are much heavier than the gravitino in either case.

%

\begin{thebibliography}{999}
%
%
\bibitem{5D_M} P.~Horava and E.~Witten, Nucl.~Phys.~\textbf{B460}, (1996) 506;
 \textbf{B475}, (1996) 94;
 A.~Lukas, B.A.~Ovrut, K.S.~Stelle and D.~Waldram, Phys.~Rev.~\textbf{D59}, (1999) 
 086001;Nucl.~Phys.~\textbf{B552}, (1999) 246. 

\bibitem{RS} L.~Randall and R.~Sundrum, Phys.~Rev.~Lett.~\textbf{83}, (1999) 3370;
 R.~Altendorfer, J.~Bagger and D.~Nemeschansky, Phys.~Rev.~\textbf{D63}, (2001) 125025;
 A.~Falkowski, Z.~Lalak and S.~Pokorski, Phys.~Lett.~\textbf{B491}, (2000) 172. 

\bibitem{AS_moduli} H.~Abe and Y.~Sakamura, {\tt arXiv:0709.3791}. 

\bibitem{KO} T.~Kugo and K.~Ohashi, Prog.~Theor.~Phys.~\textbf{105}, (2001) 323;
 \textbf{108}, (2002) 203;
 T.~Fujita and K.~Ohashi, Prog.~Theor.~Phys.~\textbf{106}, (2001) 221;
 T.~Fujita, T.~Kugo and K.~Ohashi, Prog.~Theor.~Phys.~\textbf{106}, (2001) 671. 

\bibitem{N1_offshell} F.~Paccetti Correia, M.G.~Schmidt 
 and Z.~Tavartkiladze, Nucl.~Phys.~\textbf{B709}, (2005) 141;
 H.~Abe and Y.~Sakamura, JHEP~\textbf{0410}, (2004) 013. 

\bibitem{offshell_DR} F.~Paccetti Correia, M.G.~Schmidt 
 and Z.~Tavartkiladze, Nucl.~Phys.~\textbf{B751}, (2006) 222;
 H.~Abe and Y.~Sakamura, Phys.~Rev.~\textbf{D75}, (2007) 025018. 

\bibitem{sub_AS} H.~Abe and Y.~Sakamura, Phys.~Rev.~\textbf{D71}, (2005) 105010;
 \textbf{D73}, (2006) 125013. 

\bibitem{BFbound} P.~Breitenlohner and D.Z.~Freedman, 
 Phys.~Lett.~\textbf{115B}, (1982) 197. 

\bibitem{SUSY_RS} M.A.~Luty and R.~Sundrum, Phys.~Rev.~\textbf{D64}, (2001) 
 065012. 

\bibitem{MO} N.~Maru and N.~Okada, Phys.~Rev.~\textbf{D70}, (2004) 025002. 

\bibitem{KKLT} S.~Kachru, R.~Kallosh, A.~Linde and S.P.~Trivedi, 
 Phys.~Rev.~\textbf{D68}, (2003) 046005. 

\bibitem{choi} K.~Choi, A.~Falkowski, H.P.~Nilles, M.~Olechowski 
 and S.~Pokorski, JHEP~\textbf{0411}, (2004) 076;
 K.~Choi, {\tt arXiv:0705.3330}. 

\end{thebibliography}
%

\end{document}